\documentclass[prb,superscriptaddress,twocolumn,amsmath,amssymb,nofootinbib]{revtex4-2}
\usepackage{hyperref}
\usepackage{graphicx}
\usepackage{natbib}

\usepackage[dvipsnames]{xcolor}

\usepackage{lipsum}
\newcommand{\CC}{\mathrm C}
\newcommand{\HH}{\mathrm H}
\newcommand{\LL}{\mathrm L}
\newcommand{\RR}{\mathrm R}
\newcommand{\e}{\mathrm e}
\newcommand{\kB}{k_\mathrm B}

\begin{document}

\title{Thermal control across a chain of electronic nanocavities}
\date{\today}

\author{\'Etienne Jussiau}
\email{ejussiau@ur.rochester.edu}
\affiliation{Department of Physics and Astronomy, University of Rochester, Rochester, NY 14627, USA}
\affiliation{Center for Coherence and Quantum Optics, University of Rochester, Rochester, NY 14627, USA}
\author{Sreenath K. Manikandan}
\email{skizhakk@ur.rochester.edu}
\affiliation{Department of Physics and Astronomy, University of Rochester, Rochester, NY 14627, USA}
\affiliation{Center for Coherence and Quantum Optics, University of Rochester, Rochester, NY 14627, USA}
\author{Bibek Bhandari}
\email{bbhandar@ur.rochester.edu}
\affiliation{Department of Physics and Astronomy, University of Rochester, Rochester, NY 14627, USA}
\affiliation{Center for Coherence and Quantum Optics, University of Rochester, Rochester, NY 14627, USA}
\affiliation{NEST, Scuola Normale Superiore and Istituto Nanoscienze-CNR, I-56126 Pisa, Italy}
\author{Andrew N. Jordan}
\email{jordan@pas.rochester.edu}
\affiliation{Department of Physics and Astronomy, University of Rochester, Rochester, NY 14627, USA}
\affiliation{Center for Coherence and Quantum Optics, University of Rochester, Rochester, NY 14627, USA}
\affiliation{Institute for Quantum Studies, Chapman University, Orange, CA 92866, USA}

\begin{abstract}
We study a chain of alternating hot and cold electronic nanocavities---connected to one another via resonant-tunneling quantum dots---with the intent of achieving precise thermal control across the chain. This is accomplished by positioning the dots' energy levels such that a predetermined distribution of heat currents is realized across the chain in the steady state. The number of electrons in each cavity is conserved in the steady state which constrains the cavities' chemical potentials. We determine these chemical potentials analytically in the linear response regime where the energy differences between the dots' resonant levels and the neighboring chemical potentials are much smaller than the thermal energy. In this regime, the thermal control problem can be solved exactly, while, in the general case, thermal control can only be achieved in a relative sense, that is, when one only preassigns the ratios between different heat currents.
We apply our results to two different cases: We first demonstrate that a ``heat switch'' can be easily realized with three coupled cavities, and then we show that our linear response results can provide accurate results in situations with a large number of cavities.
\end{abstract}
\maketitle

\section{Introduction}

The remarkable progress made in the last few decades in developing and demonstrating control over various quantum technology platforms---including superconducting qubits~\cite{murch2013observing,kjaergaard2020superconducting,vijay2012stabilizing,weber2014mapping}, semiconductor spin qubits~\cite{de2013ultrafast,awschalom2013quantum,qiao2020conditional}, and ion traps~\cite{haffner2008quantum,schneider2012experimental}---invites considerable interest in precise thermodynamic characterization of various energy needs when many quantum devices are operating collectively, and in understanding the fundamental limits imposed by the laws of thermodynamics on the operation of such devices~\cite{wang2002,whitney2018quantum,pekola2015towards,Koski2014,olivier2019}. The field of thermodynamics in the quantum regime has developed considerably over the past years to address this and related questions of contemporary interest~\cite{benenti2017fundamental,sothmann2014thermoelectric,giazotto2006opportunities,manikandan2019superconducting,ronzani2018}. In almost every situation where a delicate quantum technology platform is put in contact with thermal resources---which is also the canonical premise of quantum thermodynamics---we observe that thermal effects take over, and ultimately, the behavior of the system in the long-time limit is dictated by thermodynamic principles. This still permits practical quantum device applications such as steady-state heat engines and refrigerators at the nanoscale~\cite{jordan2013powerful,sothmann2014thermoelectric,sothmann2012rectification,sothmann2013powerful,sanchez2013correlations,jaliel2019experimental}. In these nanoscale devices, whether the device acts as a heat engine or a refrigerator depends on the choice of parameters~\cite{Campisi2020,sothmann2014thermoelectric,manikandan2020autonomous}. Although management of heat across a single nanoscale device has been well studied~\cite{benenti2017fundamental,sothmann2012rectification,sothmann2013powerful,sothmann2014thermoelectric,manikandan2020autonomous,fornieri2017,timossi2018,Li2012,bosisio2015}, an on-demand heat flow across a chain of quantum units is still an unsolved problem. The controlled flow of heat in quantum circuits is highly desired. This is because overheating can lead to the degradation of performance or damage to the system.

Here we address a closely related problem of great significance to quantum technology at the nanoscale, which we henceforth refer to as \textit{thermal control}; in particular, we ask whether it is possible to preassign values for heat currents in a chain of electronic nanocavities consisting of alternating hot and cold regions. Our goal is to realize a predetermined distribution of heat currents from arbitrary initial conditions by means of minimal external quantum control. We present a scheme where by suitably positioning the resonant energies of the quantum dots connecting the electronic nanocavities together (see Fig.~\ref{fig:sketch}), we are able to achieve thermal control such as defined above. Dot energy positions can be controlled by independent gate voltages, which renders our prescription particularly simple to implement experimentally.

Such precise control of heat currents is indeed a difficult problem, and analytic solutions are particularly challenging. The difficulty is compounded when the number of constituting units becomes very large. Yet, we find that a remarkably simple solution exists when the energy differences between the dot energies and the chemical potentials of the cavities they connect are small compared to the thermal energy. More specifically, the configuration of dot energies which achieves a preassigned distribution of heat currents across a chain of electronic nanocavities can be estimated analytically.

A major advantage of our proposal is that it allows the extraction of heat from (and therefore refrigeration of) any number of electronic nanocavities across the chain, at the expense of heating up the remaining nanocavities. This is of great interest for various quantum technology platforms that exist in superconducting as well as semiconductor based spin qubit platforms; one could imagine the cavities that are cooled down are coupled to qubits which repeatedly undergo cycles of irreversible computation, releasing a $\kB T\ln2$ amount of heat per cycle as a result (determined by the Landauer bound~\cite{landauer1961irreversibility}), as depicted in Fig.~\ref{fig:sketch}. The remaining units where the excess heat is dumped practically act like exhaust pipes or intermittent drains for wasted energy in the circuit, offering an elegant scheme for on-chip management of excess heat. A good analogy can be made to an advanced air-conditioning system for workplaces, where heat can be extracted from multiple locations (potentially maintained at different preassigned ambient thermostat temperatures), and perfectly channeled out through the exhausts in a centralized fashion.


This article is organized as follows: In Sec.~\ref{Sec-Model}, we characterize thermoelectric transport across the chain of nanocavities considered here, and we present the conservation laws which determine the cavities' chemical potentials. In Sec.~\ref{Sec-LinearResponse}, we show that these chemical potentials can be calculated analytically in the linear response regime where the energy differences between a dot's resonant energy and the chemical potentials of the neighboring cavities are assumed to be small. We use this solution to solve the thermal control problem in Sec.~\ref{Sec-ThCtrl}. The absolute thermal control problem, where the values of all heat currents are preassigned, can only be solved in the linear response regime, but we show that relative thermal control, where only the ratios between different heat currents are preassigned, can be achieved in the general case. Finally, we present two examples illustrating the versatility of our scheme in Sec.~\ref{Sec-Examples}.


\section{Model and formulation}
\label{Sec-Model}

\begin{figure}
	\centering
	\includegraphics[width=\linewidth]{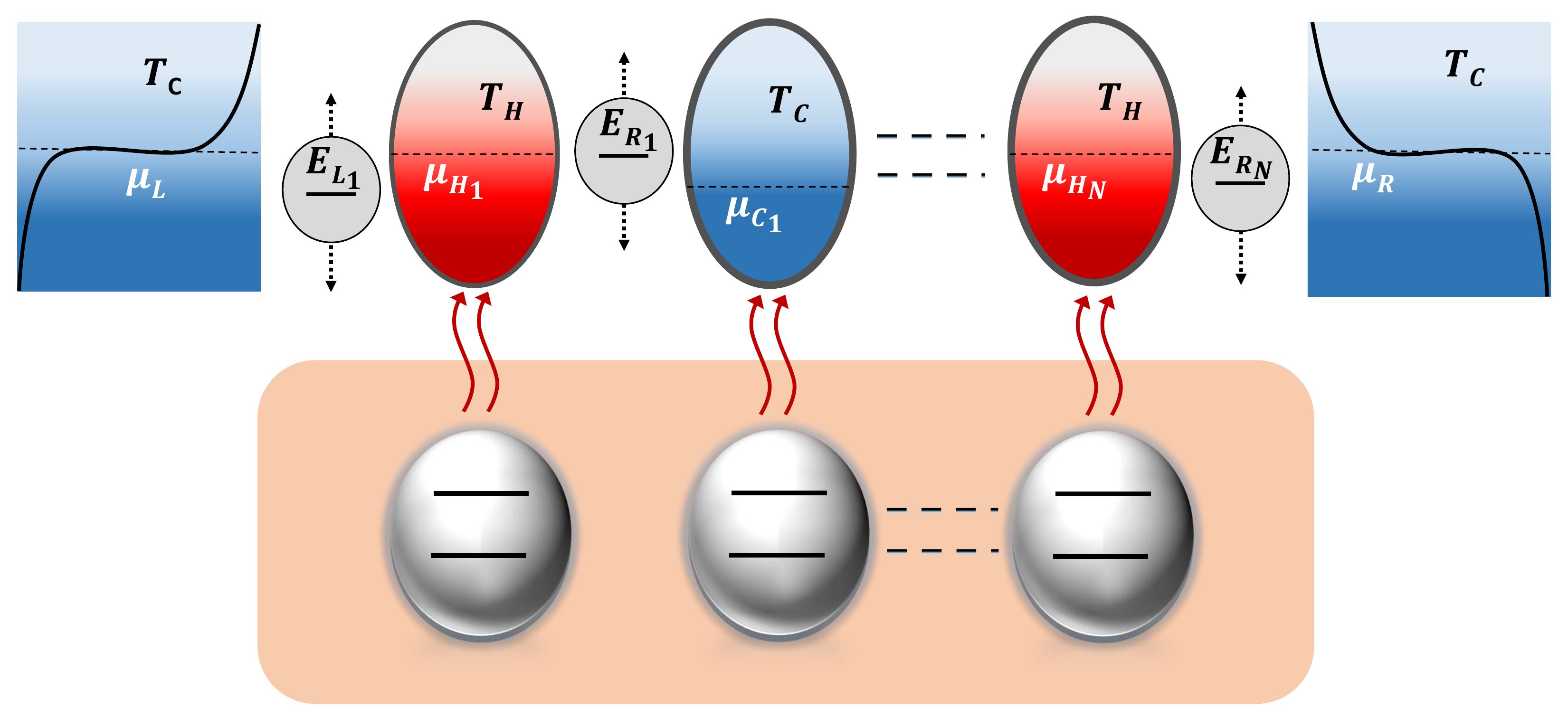}
	\caption{A possible realization of thermal control across a chain of electronic nanocavities. We consider a series of electronic nanocavities connected to external reservoirs and coupled to each other via quantum dots with prescribed resonant energies. The cavities are exchanging energy in the form of heat with ancillary quantum systems, possibly a bunch of qubits undergoing an irreversible cycle of computation and dissipating energy in the form of heat. Our objective is to find the dot energy configuration such that a preassigned distribution of heat currents across the nanocavities is realized in the steady state. These positions for the quantum dots' energies are determined once the target distribution for heat currents across the chain is specified.}
	\label{fig:sketch}
\end{figure}

We consider a chain of an arbitrary number of alternating hot and cold nanocavities, with respective temperatures~$T_\HH$ and~$T_\CC$, which are connected to each other via resonant-tunneling quantum dots (see Fig.~\ref{fig:sketch}). In addition, the extremities of the chain are electron reservoirs (leads) whose temperatures and chemical potentials can be controlled externally. The temperature differences all along the chain give rise to a flow of electrons, and an associated flow of heat which will be the main focus of this article. We assume that each cavity in the chain is coupled to an ancillary quantum system---represented by qubits in Fig.~\ref{fig:sketch}---which acts as a heat reservoir and maintains the cavity's temperature constant through energy exchange. However, the cavity does not exchange particles with the ancillary system, so its chemical potential is constrained by the conservation of the number of electrons it contains in the steady state. Furthermore, we assume that electrons quickly thermalize upon entering a cavity due to strong electron-electron and electron-phonon inelastic scattering processes.
As a consequence, the electron population in a cavity follows the Fermi-Dirac distribution, $f(E-\mu,T)=(1+\e^{(E-\mu)/\kB T})^{-1}$. This approximation is valid provided that the thermalization time for electrons in a cavity is much shorter compared to the typical time it will spend there before hopping to an adjacent quantum dot. We must emphasize here that electronic nanocavities and quantum dots are vastly different objects; namely, cavities are much bigger. As a consequence, a cavity typically contains a macroscopic number of electrons, and addition or removal of an electron from a cavity will then have a negligible impact on the many-body state inside it. The situation is different for quantum dots which are small enough that a single resonant energy level appears due to electron confinement. Coulomb repulsion will prevent such a level from being doubly occupied. Thus, electrons on a resonant-tunneling quantum dot are effectively noninteracting, which justifies the use of the Landauer-Büttiker formalism in Eq.~\eqref{LBformula} below.

A similar system comprising a single cavity has already been extensively studied, primarily with the intent of cooling down the cold central cavity~\cite{edwards1993quantum,edwards1995cryogenic,prance2009electronic}, but, in the case of a hot cavity, the device can also operate as an energy harvester~\cite{jordan2013powerful,jaliel2019experimental} or an absorption refrigerator~\cite{manikandan2020autonomous}. In this last case, it has been shown that a rectified heat current appears when the two reservoirs at the extremities are held at the same temperature. Here, we make use of this property to drive a heat current across a chain with multiple cavities.  We then assume that the leads at the extremities of the chain are held at the same temperature~$T_\CC$ (equal to that of the cold cavities) and chemical potential $\mu$. Without loss of generality, we set the zero of energy at $\mu$ hereafter, $\mu=0$. Reference~\cite{Humphrey} considers an analogous setup: A chain of electronic reservoirs with alternating temperatures, connected to each other via quantum dots, is used to emulate the behavior of an adiabatically rocked electron ratchet. In this situation, it is assumed that all the chemical potentials across the chain can be controlled externally by bias voltages. This is a crucial difference with the present analysis: The nanocavities considered here do not exchange particles with the environment and their chemical potentials are then constrained by particle conservation. As a consequence, it is necessary to infer these chemical potentials in order to fully characterize the system. This is a particularly challenging task in the general case.

A hot cavity, with temperature~$T_\HH$, at position~$k$ within the chain is denoted by $\HH_k$; it is connected to two cold cavities, with temperature~$T_\CC$, cavity~$\CC_{k-1}$ on the left via dot $E_{\LL_k}$, and cavity~$\CC_k$ on the right via dot~$E_{\RR_k}$. The reservoirs at the extremities of the chain are denoted by $\LL=\CC_0$ and $\RR=\CC_N$, where $N$ is the number of hot cavities. The particle current flowing from site~$\kappa$ in the chain to an adjacent site~$\kappa'$ is given by the Landauer-B\"uttiker formula~\cite{Landauer1957,Landauer1970,Buttiker1986,buttiker1988coherent,benenti2017fundamental},
\begin{equation}
        j_{\kappa\to\kappa'}=\frac2h\int\mathrm dE\,\mathcal T_{\kappa\kappa'}(E)(f_\kappa(E)-f_{\kappa'}(E)),
        \label{LBformula}
\end{equation}
where $f_\kappa(E)=f(E-\mu_\kappa,T_\kappa)$, $\mathcal T_{\kappa\kappa'}(E)$ is the transmission function for the quantum dot connecting the two sites, and $h=2\pi\hbar$ is Planck's constant. In the resonant-tunneling regime, the transmission function reads~\cite{buttiker1988coherent,benenti2017fundamental}
\begin{equation}
    \mathcal T_{\kappa\kappa'}(E)=\frac{\gamma^2}{(E-E_{\kappa\kappa'})^2+\gamma^2},
    \label{T_fct}
\end{equation}
where $E_{\kappa\kappa'}$ is the resonant energy of the dot ($E_{\kappa\kappa'}=E_{\LL_k}$ or $E_{\kappa\kappa'}=E_{\RR_k}$) and $\gamma$ is the level width. Here we have assumed for simplicity that the dot is symmetrically coupled to~$\kappa$ and~$\kappa'$, and that this coupling does not depend on its position in the chain.

The chemical potentials of the $N$ hot cavities and $N-1$ cold ones within the chain are constrained by the conservation of the number of electrons in each of these cavities. They are thus determined by solving the following system of $2N-1$ coupled equations for the chemical potentials~$\mu_{\HH_k}$ and~$\mu_{\CC_k}$,
\begin{equation}
\begin{cases}
    j_{\CC_{k-1}\to\HH_k}+j_{\CC_k\to\HH_k}=0&\text{for $k=1,\dots,N$}\\
    j_{\HH_k\to\CC_k}+j_{\HH_{k+1}\to\CC_k}=0&\text{for $k=1,\dots,N-1$}.
\end{cases}
\label{sys_e-}
\end{equation}
Hereafter, we denote by $j$ the particle current flowing across the chain (from left to right). For all $k$, it satisfies
\begin{equation}
    j=j_{\CC_{k-1}\to\HH_k}=-j_{\HH_k\to\CC_{k-1}}=j_{\HH_k\to\CC_k}=-j_{\CC_k\to\HH_k}.
\end{equation}

Furthermore, the flow of electrons through the dot connecting sites~$\kappa$ and~$\kappa'$ carries an energy current
\begin{equation}
    J_{\kappa\to\kappa'}=\frac2h\int\mathrm dE\,E\mathcal T_{\kappa\kappa'}(E)(f_\kappa(E)-f_{\kappa'}(E)).
\end{equation}
The energy in each cavity is also conserved; however, the heat currents flowing from the ancillary systems must be taken into account here. We then have
\begin{equation}
\begin{cases}
    J_{\CC_{k-1}\to\HH_k}+J_{\CC_k\to\HH_k}+\dot Q_{\HH_k}=0&\text{for $k=1,..,N$}\\
    J_{\HH_k\to\CC_k}+J_{\HH_{k+1}\to\CC_k}+\dot Q_{\CC_k}=0&\text{for $k=1,..,N-1$},
\end{cases}
\label{sys_NRG}
\end{equation}
where $\dot Q_\kappa$ denotes the heat current into cavity~$\kappa$ that maintains its temperature constant.
Note that, since the net particle current out of each cavity is zero, the corresponding heat and energy currents coincide. This is also the case for the leads at the extremities of the chain because we have taken their common chemical potential to be the zero of energies, even though the currents flowing out of these leads do not vanish: $j_{\LL\to\HH_1}=j_{\HH_N\to\RR}=j$.

The conservation laws expressed in Eqs.~\eqref{sys_e-} and~\eqref{sys_NRG} can be dramatically simplified in the sequential tunneling regime where it is assumed that dots are weakly coupled to the neighboring sites such that their level width is much smaller than the typical thermal energy, $\gamma\ll\kB T_\CC$. In this limit, the transmission function in Eq.~\eqref{T_fct} becomes infinitely sharp such that only electrons whose energy exactly matches the dot's resonant energy can go through:
$\mathcal T_{\kappa\kappa'}(E)\simeq\pi\gamma\delta(E-E_{\kappa\kappa'})$. The particle and energy currents then become
\begin{align}
&j_{\kappa\to\kappa'}=\frac\gamma\hbar\big(f_\kappa(E_{\kappa\kappa'})-f_{\kappa'}(E_{\kappa\kappa'})\big),\label{j_WeakCoupling}\\
&J_{\kappa\to\kappa'}=\frac{\gamma E_{\kappa\kappa'}}\hbar\big(f_\kappa(E_{\kappa\kappa'})-f_{\kappa'}(E_{\kappa\kappa'})\big).
\end{align}
Since all transmitted electrons have the same energies, the energy current is directly proportional to the particle current, $J_{\kappa\to\kappa'}=E_{\kappa\kappa'}j_{\kappa\to\kappa'}$. This tight coupling of
the particle and energy flows dramatically simplifies the calculation of the heat currents~$\dot Q_\kappa$ as Eq.~\eqref{sys_NRG} now reads
\begin{equation}
\begin{cases}
    \dot Q_{\HH_k}=j(E_{\RR_k}-E_{\LL_k}),&1\le k\le N\\
    \dot Q_{\CC_k}=j(E_{\LL_{k+1}}-E_{\RR_k}),&1\le k<N.
\end{cases}
\label{sys_NRG-0}
\end{equation}
We then understand that the particle current across the chain~$j$ is the central quantity in our problem since all the thermoelectric properties of the system can be deduced from it. To obtain an analytical expression for this current, it is necessary to solve the system in Eq.~\eqref{sys_e-} for the cavities' chemical potentials. For narrow level widths, this system of equations reads
\begin{equation}
\begin{cases}
\begin{aligned}[b]
&f_{\CC_{k-1}}(E_{\LL_k})-f_{\HH_k}(E_{\LL_k})\\
&+f_{\CC_k}(E_{\RR_k})-f_{\HH_k}(E_{\RR_k})=0,
\end{aligned}&1\le k\le N\\
\\
\begin{aligned}[b]
&f_{\HH_k}(E_{\RR_k})-f_{\CC_k}(E_{\RR_k})\\
&+f_{\HH_{k+1}}(E_{\LL_{k+1}})-f_{\CC_k}(E_{\LL_{k+1}})=0,
\end{aligned}&1\le k<N.
\end{cases}
\label{sys_e-0}
\end{equation}




\section{Linear response solution}
\label{Sec-LinearResponse}

The system in Eqs.~\eqref{sys_e-0} expressing particle conservation in the weak-coupling regime can only be solved exactly in the simplest case with a single cavity ($N=1$), where the system is reduced to a single equation~\cite{manikandan2020autonomous}. Nevertheless, we show in this section that a solution for an arbitrary number of cavities can be obtained, provided that each dot energy is positioned close to the chemical potentials of the cavities (or lead) it is connected to. In a similar spirit as the linear response approximation widely used to study quantum thermoelectric devices~\cite{benenti2017fundamental}, we expand the Fermi factors to first order, $f(E-\mu,T)\simeq1/2-(E-\mu)/4\kB T$, which is a valid approximation when $|E-\mu|\ll\kB T$. Within this linear response approximation, Eqs.~\eqref{sys_e-0} become
\begin{equation}
\begin{cases}
\displaystyle
    \begin{aligned}[b]
    &\frac{E_{\LL_k}-\mu_{\HH_k}}{T_\HH}-\frac{E_{\LL_k}-\mu_{\CC_{k-1}}}{T_\CC}\\
    &+\frac{E_{\RR_k}-\mu_{\HH_k}}{T_\HH}-\frac{E_{\RR_k}-\mu_{\CC_k}}{T_\CC}=0,
    \end{aligned}&1\le k\le N\\
    \\
  \begin{aligned}[b]
  &\frac{E_{\RR_k}-\mu_{\CC_k}}{T_\CC}-\frac{E_{\RR_k}-\mu_{\HH_k}}{T_\HH}\\
  &+\frac{E_{\LL_{k+1}}-\mu_{\CC_k}}{T_\CC}-\frac{E_{\LL_{k+1}}-\mu_{\HH_{k+1}}}{T_\HH}=0,
  \end{aligned}&1\le k<N.
\end{cases}
\label{sys_Lin}
\end{equation}

This system can be solved analytically without further assumption and we find\footnote{A detailed derivation is presented in Appendix~\ref{app_ChemPot}.} 
\begin{subequations}
\begin{align}
    &\mu_{\HH_k}=\left(\frac{T_\HH}{T_\CC}-1\right)\left(\sum_{l=1}^{k-1}E_{\RR_l}-\sum_{l=1}^kE_{\LL_l}-\frac{(2k-1)\Delta}{2N}\right),\\
    &\mu_{\CC_k}=\left(1-\frac{T_\CC}{T_\HH}\right)\left(\sum_{l=1}^k(E_{\RR_l}-E_{\LL_l})-\frac {k\Delta}N\right),
\end{align}
\label{muHCk}%
\end{subequations}
where we have denoted $\Delta=\sum_{k=1}^N(E_{\RR_k}-E_{\LL_k})$. Typically, the above solution is a relevant approximation when the energy differences between the dots' resonant levels and the chemical potential of the reservoirs at the extremities of the chain are much smaller than the thermal energy. Since we consider $\mu_\LL=\mu_\RR=0$, this condition reduces to $|E_{\LL_k}|,|E_{\RR_k}|\ll\kB T_\CC$.

These chemical potentials enable us to obtain the particle current across the chain,
\begin{equation}
j=\frac{\gamma\Delta}{8N\hbar\kB\Theta},
\label{j_LinearResponse}
\end{equation}
where $\Theta=1/(T_\CC^{-1}-T_\HH^{-1})$. It is interesting to notice that the current across the chain depends on the individual positions of the dot energies through $\Delta$.

The heat currents are straightforwardly deduced from Eq.~\eqref{j_LinearResponse}:
\begin{subequations}
\begin{align}
    &\dot Q_{\HH_k}=\frac{\gamma\Delta(E_{\RR_k}-E_{\LL_k})}{8N\hbar\kB\Theta},\\
    &\dot Q_{\CC_k}=\frac{\gamma\Delta(E_{\LL_{k+1}}-E_{\RR_k})}{8N\hbar\kB\Theta}.
\end{align}
\label{QHCk}%
\end{subequations}
One should note, however, that the heat currents at the extremities of the chain must be expressed differently since both leads are connected only to a single dot:
\begin{subequations}
\begin{align}
    &\dot Q_\LL=J_{\LL\to\HH_1}=\frac{\gamma\Delta E_{\LL_1}}{8N\hbar\kB\Theta},\label{QL}\\
    &\dot Q_\RR=J_{\RR\to\HH_N}=-\frac{\gamma\Delta E_{\RR_N}}{8N\hbar\kB\Theta}.\label{QR}
\end{align}
\label{QLR}%
\end{subequations}

\section{Thermal control across the chain}
\label{Sec-ThCtrl}

\subsection{Linear response solution}

So far, we have considered that the positions of dot energies across the chain were known, and derived the corresponding heat current distribution across the chain given this input. In this section, we turn our perspective around and consider dot energy positions as free parameters that can be tuned so that heat currents across the chain match a preassigned distribution. This problem of spatial management of heat within the device is referred to as thermal control. The thermal control problem is essentially expressed by Eq.~\eqref{QHCk}, along with the boundary conditions given in Eq.~\eqref{QLR}. This forms a system a coupled equations which must be solved for the dot energies $E_{\LL_k},E_{\RR_k}$. The thermal control problem features $2N+1$ heat currents in total ($\dot Q_{\HH_k}$ for $k=1,\dots,N$, $\dot Q_{\CC_k}$ for $k=1,\dots,N-1$, $\dot Q_\LL$ and $\dot Q_\RR$) but only $2N$ dot energies that can be tuned ($E_{\LL_k},E_{\RR_k}$ for $k=1,\dots,N$), so the system seems overdetermined. Actually, the laws of thermodynamics impose further constraints on the thermal distributions that can be considered for thermal control. In the steady state, the first law reads
\begin{equation}
   \dot Q_\LL+\dot Q_\HH+\dot Q_\CC+\dot Q_\RR=0,
    \label{1stLaw}
\end{equation}
where $\dot Q_\HH=\sum_{k=1}^N\dot Q_{\HH_k}$ and $\dot Q_\CC=\sum_{k=1}^{N-1}\dot Q_{\CC_k}$. This means that there are only $2N$ independent heat currents in the problem. Furthermore, the second law can be expressed as follows:
\begin{equation}
    \frac{\dot Q_\LL+\dot Q_\CC+\dot Q_\RR}{T_\CC}+\frac{\dot Q_\HH}{T_\HH}\le0.
    \label{2ndLaw}
\end{equation}
Taking Eq.~\eqref{1stLaw} into account, Eq.~\eqref{2ndLaw} can be rewritten as $\dot Q_\HH(T_\CC^{-1}-T_\HH^{-1})\ge0$, which implies $\dot Q_\HH\ge0$, since $T_\HH>T_\CC$. This is merely a rephrasing of Clausius's celebrated statement~\cite{Clausius}:  ``Heat can never pass from a colder to a warmer body without some other change, connected therewith, occurring at the same time.''

Making use of these properties, we obtain the solution to the thermal control problem,\footnote{A detailed derivation is presented in Appendix~\ref{app_ThCtrl}.}
\begin{subequations}
\begin{align}
    &E_{\LL_k}=\pm\sqrt{\frac{8\hbar\kB\Theta}{\gamma\langle\dot Q_\HH\rangle}}\left(\dot Q_\LL+\sum_{l=1}^{k-1}(\dot Q_{\HH_l}+\dot Q_{\CC_l})\right),\label{EL_ThCtrl}\\
    &E_{\RR_k}=\pm\sqrt{\frac{8\hbar\kB\Theta}{\gamma\langle\dot Q_\HH\rangle}}\left(\dot Q_\LL+\sum_{l=1}^k\dot Q_{\HH_l}+\sum_{l=1}^{k-1}\dot Q_{\CC_l}\right),
\end{align}
\label{ThCtrl_Solution}%
\end{subequations}
where $\langle\dot Q_\HH\rangle=\dot Q_\HH/N$ is the average heat current extracted from hot cavities. The $\pm$ sign above expresses the fact that there are two distinct solutions to the problem. This is a consequence of particle-hole symmetry in the system: the $+$~sign corresponds to thermal transport mediated by hot electrons, while the $-$~sign corresponds to thermal transport mediated by hot holes, similarly to what has been observed in Ref.~\cite{manikandan2020autonomous}.

Interestingly, we notice that the steady-state entropy production rate~$\dot S=\dot Q_\HH/\Theta$ appears in the denominator in Eqs.~\eqref{ThCtrl_Solution}. However, one should not conclude that it is necessary to consider a situation where $\dot S$ is large in order to have $E_{\LL_k},E_{\RR_k}\ll\kB T_\CC$ so that the linear response approximation holds. On the contrary, we find that our solution to the thermal control problem is consistent with the linear response approximation when entropy production is limited because a large value for $\dot S$ typically implies a large value for the numerator in Eqs.~\eqref{ThCtrl_Solution}.



\subsection{Accuracy of the thermal control solution}

The solution to the thermal control problem given in Eqs.~\eqref{ThCtrl_Solution} above has been derived assuming that the linear response regime could be considered. Now that we have explicit expressions for all dot energies within the chain, it is necessary to verify that these are indeed consistent with the linear response approximation. Hence, we must infer general conditions on the heat current distributions under which the dot energies in Eqs.~\eqref{ThCtrl_Solution}, as well as the chemical potentials in Eqs.~\eqref{muHCk}, satisfy the linear response approximation. Such conditions will basically determine the range of validity of our thermal control solution.

Typically, we find that the thermal control solution in Eqs.~\eqref{ThCtrl_Solution} yields accurate results when the preassigned thermal distributions satisfy\footnote{A detailed derivation is presented in Appendix~\ref{app_ValidityThCtrl}.}
\begin{subequations}
\begin{align}
    &\langle\dot Q_\HH\rangle\ll\frac{\gamma\kB T_\CC^2}{\hbar\Theta},\label{ThCtrl_Validity-a}\\
    &\frac{\delta^2}{\langle\dot Q_\HH\rangle}\ll\frac{\gamma\kB T_\CC^2}{\hbar\Theta},\label{ThCtrl_Validity-b}\\
    &\frac{\sigma^2}{\langle\dot Q_\HH\rangle}\ll\frac{\gamma\kB\Theta}\hbar,\label{ThCtrl_Validity-c}
\end{align}
\label{ThCtrl_Validity}%
\end{subequations}
where $\delta^2$ is the ``distance'' between the thermal distributions for hot and cold cavities, and $\sigma^2$ is the former's variance,
\begin{align}
    &\delta^2=\frac1N\sum_{k=1}^N\left(\dot Q_{\CC_{k-1}}+\dot Q_{\HH_k}\right)^2,\\
    &\sigma^2=\frac1N\sum_{k=1}^N\left(\dot Q_{\HH_k}-\langle\dot Q_\HH\rangle\right)^2.
\end{align}

According to Eq.~\eqref{ThCtrl_Validity-a}, the first condition to obtain reliable results using the thermal control solution in Eqs.~\eqref{ThCtrl_Solution} is to consider thermal distributions featuring relatively small heat currents on average. Such a condition is typical of the linear response approximation which is not compatible with substantial particle or energy flows. As alluded to earlier, Eq.~\eqref{ThCtrl_Validity-a} also introduces a limitation on entropy production, namely, $\dot S\ll N\gamma\kB(T_\CC/\Theta)^2/\hbar$. This indicates that our solution should be more accurate for configurations close to reversibility.

Furthermore, Eq.~\eqref{ThCtrl_Validity-b} implies that the thermal distributions for hot and cold cavities should be close to being mutually opposite. In such a case, the dot energies given in Eqs.~\eqref{ThCtrl_Solution} will be small since heat currents from hot and cold cavities compensate one another. 

Finally, the quantity~$\sigma^2/\langle\dot Q_\HH\rangle$ appearing at the right-hand side of Eq.~\eqref{ThCtrl_Validity-c} is the dispersion index for the distribution of heat currents extracted from hot cavities; its smallness then corresponds to thermal distributions whose values do not spread far from the average value~$\langle\dot Q_\HH\rangle$. Since $T_\CC<\Theta$, the bounds on $\langle\dot Q_\HH\rangle$ and $\delta^2/\langle\dot Q_\HH\rangle$ are tighter than that on $\sigma^2/\langle\dot Q_\HH\rangle$.

Interestingly, we observe that the conditions obtained in Eqs.~\eqref{ThCtrl_Validity} only involve the temperatures~$T_\CC$ and~$T_\HH$ through the factors $\Theta$ and $T_\CC^2/\Theta$. On the one hand, when $T_\CC$ is fixed, $\Theta$ decreases with $T_\HH$, which implies that $T_\CC^2/\Theta$ increases. On the other hand, when $T_\HH$ is fixed, $\Theta$ increases with $T_\CC$, while $T_\CC^2/\Theta$ first increases and then decreases after having reaches a maximum for $T_\CC=T_\HH/2$. Note, however, that arbitrarily high values for $\Theta$ and $T_\CC^2/\Theta$ can be reached if both $T_\CC$ and $T_\HH$ are increased. This means that the thermal control solution in Eqs.~\eqref{ThCtrl_Solution} can in principle yield accurate results for any preassigned heat current distribution, provided that temperatures are appropriately chosen.


\subsection{Beyond linear response: relative thermal control}

The solution to the thermal control problem given in Eqs.~\eqref{ThCtrl_Solution} has been derived within the linear response regime, and is then valid provided that the considered thermal distribution satisfies the conditions in Eqs.~\eqref{ThCtrl_Validity}. If this is not the case, thermal control can still be realized, albeit in a weaker, relative sense. Indeed, we have assumed until now that each heat current had its precise value preassigned. We can imagine a different situation where we are only interested in controlling the fraction of the total heat allocated to each site in the chain. In this context, only the values for the ratios~$r_\kappa=\dot Q_\kappa/\dot Q_\HH$ are predetermined, without imposing any constraint on the total heat provided by the hot cavities, $\dot Q_\HH$. This indetermination on the value of $\dot Q_\HH$ removes one constraint to the problem. This means that the relative thermal control problem is underconstrained, and can therefore admit infinitely many solutions.

From Eqs.~\eqref{sys_NRG-0}, it is clear that the total heat current is given by $\dot Q_\HH=j\Delta$. Using the notation~$r_\kappa=\dot Q_\kappa/\dot Q_\HH$, the relative thermal control problem can then be expressed as follows:
\begin{equation}
\begin{cases}
    E_{\RR_k}-E_{\LL_k}=r_{\HH_k}\Delta,&1\le k\le N\\
    E_{\LL_{k+1}}-E_{\RR_k}=r_{\CC_k}\Delta,&1\le k<N,
\end{cases}
\label{sys_RelativeThCtrl}
\end{equation}
along with the boundary conditions~$E_{\LL_1}=r_\LL\Delta$ and~$E_{\RR_N}=-r_\RR\Delta$. The redundancy in Eqs.~\eqref{sys_RelativeThCtrl} lies in the fact that summing of the first line of the system for $k$ from $1$ to $N$ yields $\Delta=\sum_{k=1}^N(E_{\RR_k}-E_{\LL_k})$ since, by definition, $\sum_{k=1}^Nr_{\HH_k}=1$.

We find that the relative thermal control problem is solved by any energy configuration satisfying
\begin{subequations}
\begin{align}
    &E_{\LL_k}=\left(r_\LL+\sum_{l=1}^{k-1}(r_{\HH_l}+r_{\CC_l})\right)\Delta,\\
    &E_{\RR_k}=\left(r_\LL+\sum_{l=1}^kr_{\HH_l}+\sum_{l=1}^{k-1}r_{\CC_l}\right)\Delta,
\end{align}
\end{subequations}
where the value of $\Delta$ can be chosen arbitrarily here as one can verify that $\sum_{k=1}^N(E_{\RR_k}-E_{\LL_k})=\Delta$ is automatically ensured.


\section{Two illustrative examples}
\label{Sec-Examples}

\subsection{Various thermal control schemes in a simple case}

\begin{figure}
	\centering
	\includegraphics[width=\linewidth]{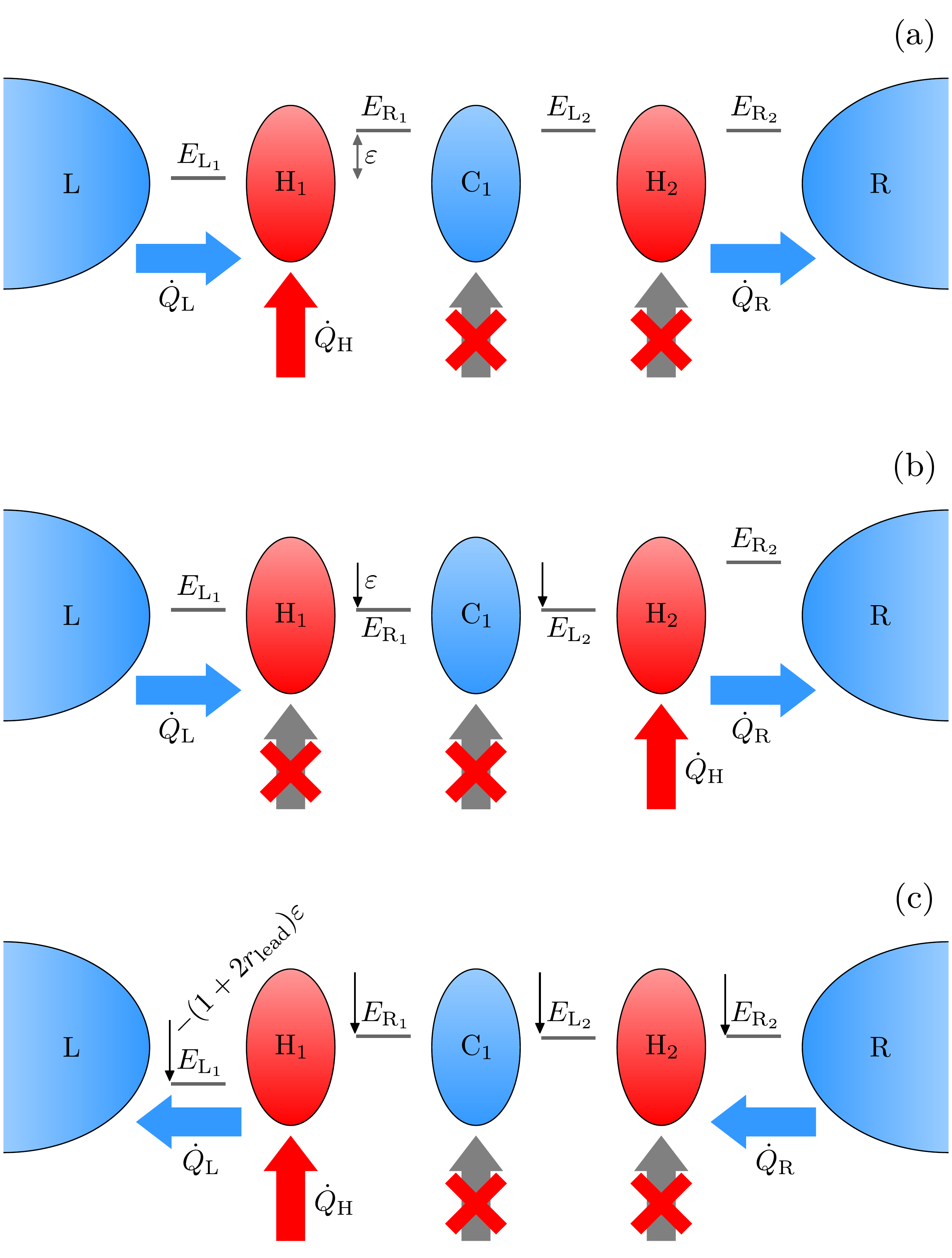}
	\caption{(a) The situation at stake here: a chain comprising two hot cavities ($\HH_1$ and $\HH_2$) and a cold one ($\CC_1$), with cold electron reservoirs ($\LL$ and $\RR$) at the extremities. We assume that $\dot Q_{\CC_1}=0$, and the other heat currents are given by $\dot Q_\LL=r_\mathrm{lead}\dot Q_\HH$, $\dot Q_{\HH_1}=r_\mathrm{cav}\dot Q_{\HH}$, $\dot Q_{\HH_2}=(1-r_\mathrm{cav})\dot Q_\HH$, and $\dot Q_\RR=-(1+r_\mathrm{lead})\dot Q_\HH$. For simplicity, the figure depicts the case~$r_\mathrm{cav}=1$. (b) By shifting the two inner dot energies~$E_{\RR_1}$ and~$E_{\LL_2}$ by the same amount~$(1-2r_\mathrm{cav})\varepsilon$, the heat currents for the two hot cavities~$\HH_1$ and~$\HH_2$ are swapped. (c) By shifting all dot energies by the same amount~$-(1+2r_\mathrm{lead})\varepsilon$, the heat currents for the leads at the extremities of the chain are swapped.}
	\label{Fig-N2}
\end{figure}

Let us now illustrate our results considering two different examples. First we present the simple case where $N=2$, highlighting the various possibilities offered by our thermal control scheme. As shown in Fig.~\ref{Fig-N2}, we consider two hot cavities~$\HH_1$ and~$\HH_2$, coupled to the reservoirs~$\LL$ and~$\RR$ respectively. In addition, $\HH_1$ and~$\HH_2$ are coupled to each other through a cold cavity~$\CC_1$. We consider the situation where there is no net heat flow out of $\CC_1$, $\dot Q_{\CC_1}=0$, and heat is extracted from the hot cavities to drive a rectified heat current between the two leads.


We first demonstrate how thermal control enables us to choose the amount of heat extracted from each hot cavity. We denote by $r_\mathrm{lead}$ and $r_\mathrm{cav}$ the fraction of heat out of $\LL$ and $\HH_1$, respectively, that is, $\dot Q_\LL=r_\mathrm{lead}\dot Q_\HH$ and $\dot Q_{\HH_1}=r_\mathrm{cav}\dot Q_\HH$. Using Eq.~(\ref{1stLaw}) along with $\dot Q_{\CC_1}=0$, we obtain $\dot Q_{\HH_2}=(1-r_\mathrm{cav})\dot Q_\HH$ and $\dot Q_\RR=-(1+r_\mathrm{lead})\dot Q_\HH$. According to Eqs.~\eqref{ThCtrl_Solution}, the dot energies solving the thermal control problem in this case\footnote{Hereafter, we work with the thermal control solution carrying a $+$~sign, but similar conclusions could be drawn considering the opposite solution.} are
\begin{subequations}
\begin{align}
   &E_{\LL_1}=r_\mathrm{lead}\varepsilon,\\
   &E_{\RR_1}=E_{\LL_2}=(r_\mathrm{lead}+r_\mathrm{cav})\varepsilon,\\
   &E_{\RR_2}=(1+r_\mathrm{lead})\varepsilon,
\end{align}
\label{ThCtrl_N2}%
\end{subequations}
where we have used the notation $\varepsilon=4\sqrt{\hbar\kB\Theta\dot Q_\HH/\gamma}$. Such a configuration (with $r_\mathrm{cav}=1$) is depicted in Fig.~\ref{Fig-N2}(a).

If we now wish to swap the heat currents extracted from the hot cavities, that is, impose $\dot Q_{\HH_1}=(1-r_\mathrm{cav})\dot Q_\HH$ and $\dot Q_{\HH_2}=r_\mathrm{cav}\dot Q_\HH$, one needs to simply change the positions of the two inner dot energies~$E_{\RR_1}$ and~$E_{\LL_2}$. Indeed, the solution to the thermal control problem in this case reads
\begin{subequations}
\begin{align}
    &E_{\LL_1}=r_\mathrm{lead}\varepsilon,\\
    &E_{\RR_1}=E_{\LL_2}=(1+r_\mathrm{lead}-r_\mathrm{cav})\varepsilon,\\
    &E_{\RR_2}=(1+r_\mathrm{lead})\varepsilon.
\end{align}
\end{subequations}
Such a configuration (with $r_\mathrm{cav}=1$) is depicted in Fig.~\ref{Fig-N2}(b). Comparing with Eqs.~\eqref{ThCtrl_N2}, we observe that dot energies~$E_{\RR_1}$ and~$E_{\LL_2}$ should be shifted by the same amount~$(1-2r_\mathrm{cav})\varepsilon$ to swap the heat extracted from the cavities~$\HH_1$ and~$\HH_2$. This is very appealing as we could imagine a setup where the two dot energies are controlled by a single voltage source.

Using the properties described above, we can imagine different kinds of ``heat switches.'' For example, let us consider $r_\mathrm{cav}=1$ in which case all the heat is extracted from cavity~$\HH_1$, as depicted in Fig.~\ref{Fig-N2}(a). As previously observed, the heat currents from cavities~$\HH_1$ and~$\HH_2$ can be swapped by shifting dot energies~$E_{\RR_1}$ and~$E_{\LL_2}$ simultaneously. Then, the heat flow from $\HH_1$ is blocked, and all the heat necessary to power the device is extracted from $\HH_2$. This is of course only true \textit{in the steady state} as the device goes through a transient regime when the positions of dot energies are changed. In the transient regime, the detailed behavior of the device generally involves nonequilibrium features which must be dealt with using sophisticated theoretical techniques~\cite{moskalets2002floquet,sothmann2012rectification}. In order to avoid such possibly detrimental effects, the dot energy positions must be varied very slowly, such that an electron going through the device would not feel the change. No additional machinery is necessary to tackle this so-called adiabatic limit as one can simply reuse the results derived in the stationary case, albeit with time-dependent parameters~\cite{brouwer1998scattering,moskalets2002dissipation,moskalets2002floquet}. These considerations are crucial as regards thermal control as this adiabatic shift of dot energies seems to be the only way to ensure full control of heat currents throughout the swap process. For example, this would be of primary importance in a situation where $\dot Q_{\CC_1}=0$ must be true at all times.

Similarly, it is possible to reverse the flow of heat from $\LL\to\RR$ to $\RR\to\LL$, that is, swap the heat currents out of the leads. In this case, we have $\dot Q_\LL=-(1+r_\mathrm{lead})\dot Q_\HH$, $\dot Q_{\HH_1}=r_\mathrm{cav}\dot Q_\HH$, $\dot Q_{\HH_2}=(1-r_\mathrm{cav})\dot Q_\HH$, and $\dot Q_\RR=r_\mathrm{lead}\dot Q_\HH$, and the thermal control solution is given by
\begin{subequations}
\begin{align}
    &E_{\LL_1}=-(1+r_\mathrm{lead})\varepsilon,\\
    &E_{\RR_1}=E_{\LL_2}=-(1+r_\mathrm{lead}-r_\mathrm{cav})\varepsilon,\\
    &E_{\RR_2}=-r_\mathrm{lead}\varepsilon.
\end{align}
\end{subequations}
Such a configuration (with $r_\mathrm{cav}=1$) is depicted in Fig.~\ref{Fig-N2}(c). Comparing with the initial situation described in Eqs.~\eqref{ThCtrl_N2}, we observe that all dot energies have been shifted by the same amount~$-(1+2r_\mathrm{lead})\varepsilon$.

\subsection{Thermal control across a long chain}

\begin{figure}
    \includegraphics[width=\linewidth]{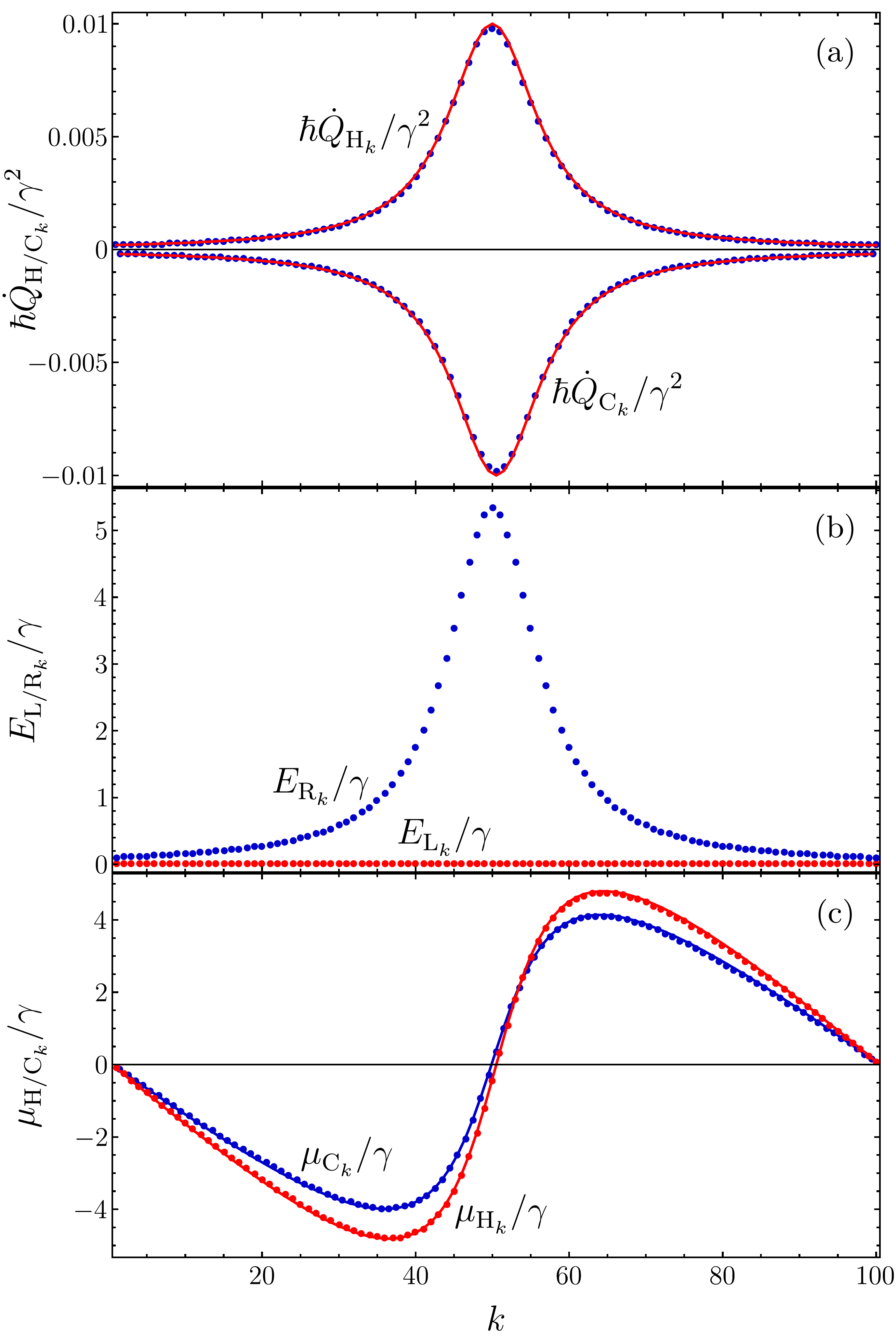}
    \caption{(a) Plots of the heat out of the hot (top) and cold (bottom) cavities across a long chain ($N=100$). The red lines correspond to the preassigned thermal distribution while the blue dots correspond to the exact result obtained solving Eqs.~\eqref{sys_e-0} numerically with the dot energies given in Eqs.~\eqref{ThCtrl_Solution} as inputs. The heat currents for the extremities~$\dot Q_\LL$ and~$\dot Q_\RR$ are not depicted in the figure. (b) The dot energies obtained with Eqs.~\eqref{ThCtrl_Solution} in this situation. The blue dots correspond to the right dot energies~$E_{\RR_k}$, and the red ones correspond to the left dot energies~$E_{\LL_k}$ which are all zero here since the thermal distributions for hot and cold cavities exactly cancel one another. (c) The corresponding chemical potentials. The solid lines correspond to the linear response result in Eqs.~\eqref{muHCk} while the dots correspond to the exact result obtained solving Eqs.~\eqref{sys_e-0} numerically with the dot energies given in Eqs.~\eqref{ThCtrl_Solution} as inputs. For this figure, we have used the thermal distributions in Eq.~\eqref{ThDist-N100} with $\sigma=7$ and $\dot Q_\mathrm{max}=0.01\gamma^2/\hbar$, while the temperatures are $T_\CC=11\gamma/\kB$ and $T_\HH=13\gamma/\kB$.}
    \label{Fig-N100}
\end{figure}

In order to illustrate the versatility of our thermal control scheme, we now analyze a radically different situation where a large number ($N=100$) of cavities are coupled together. Let us define the desired control problem:
\begin{equation}
\dot Q_{\HH/\CC_k}=\pm\frac{\sigma^2\dot Q_\mathrm{max}}{(k-n/2)^2+\sigma^2},
\label{ThDist-N100}
\end{equation}
where $\sigma$ is a dimensionless parameter characterizing the width of the thermal distribution, and $\dot Q_\mathrm{max}$ is the largest heat current in the chain (in absolute value). For simplicity, we also choose $\dot Q_\LL=0$. The solution to the absolute thermal control problem is shown in Fig.~\ref{Fig-N100}. The red lines in Fig.~\ref{Fig-N100}(a) correspond to the preassigned distributions in Eq.~\eqref{ThDist-N100}, while the blue dots represent the numerical solution to the system in Eqs.~\eqref{sys_e-0} when the dot energies are given by the thermal control solution in Eqs.~\eqref{ThCtrl_Solution}. The agreement is almost perfect: We estimate the accuracy of our prediction to be over $98\%$. For the set of parameters chosen here (see the caption of Fig.~\ref{Fig-N100}), we have $\Theta\approx72\gamma/\kB$ and $T_\CC^2/\Theta\approx1.7\gamma/\kB$ which implies that the accuracy of our approximation will be particularly impacted by the thermal distribution's average value and by the symmetry between distributions for hot and cold cavities.

\section{Conclusion}

We have investigated a chain of alternating hot and cold electronic nanocavities connected to each other via resonant quantum dots. The number of electrons in each of these cavities is conserved in the steady state which constrains their chemical potentials. This results in a system of $2N-1$ coupled equations, where $N$ is the number of hot cavities within the chain. This system cannot be solved exactly in the general case, but we have derived a solution valid in the linear response regime where the energy differences between the dots' resonant levels and the neighboring chemical potentials are small with respect to the thermal energy. In this regime, we have shown how to solve the thermal control problem: that is, finding the configuration of dot energies that would give rise to a specific preassigned distribution of heat currents across the chain. Such a solution is valid when the heat distribution we seek to produce obeys three general criteria: the heat current distributions for hot and cold cavities must be almost opposite to one another and their average and variance must be small, although the constraint on variance is somewhat looser than the other two. If this is not the case, the thermal control problem can still be solved, albeit in a weaker sense. Indeed, in this context, we are not realizing an absolute thermal distribution, but a relative one where only the fraction of the total heat allocated to each position in the chain is set beforehand.

We have illustrated our results with two different setups. In the first instance, we analyze the simple case where there are only two hot cavities in the chain ($N=2$). We show that simple operations on dot energy positions enable to swap the heat currents from given sites in the chain: Shifting two dot energies by the same amount swaps the heat extracted from the two hot cavities, and shifting all dot energies by the same amount swaps the heat currents out of the two reservoirs at the extremities of the chain. These features could be used to design a thermal switch, where dot energy positions must be changed adiabatically to ensure full thermal control at all times. Furthermore, we show that the theoretical framework developed here can also produce reliable results in a situation with a large number of cavities ($N=100$). This is a clear illustration of the versatility of our results which we envision could prove useful in a wide variety of setups. We have focused on the case of a chain of coupled cavities in the present article, but we can imagine studying more complex networks using the theoretical framework developed here which would open up new possibilities for thermal management in nanoscale systems.

\begin{acknowledgments}
This work was supported by the U.S. Department of Energy (DOE), Office of Science, Basic Energy Sciences (BES), under Award No. DE-SC0017890.
\end{acknowledgments}

\appendix

\section{Chemical potentials in the linear response regime}
\label{app_ChemPot}

The chemical potentials~$\mu_{\HH_k}$ and~$\mu_{\CC_k}$ are constrained by the conservation of the number of electrons in each cavity. In the linear response regime, this expressed by the system in Eqs.~\eqref{sys_Lin},
\begin{equation}
\begin{cases}
\displaystyle
    \begin{aligned}[b]
    &\frac{E_{\LL_k}-\mu_{\HH_k}}{T_\HH}-\frac{E_{\LL_k}-\mu_{\CC_{k-1}}}{T_\CC}\\
    &+\frac{E_{\RR_k}-\mu_{\HH_k}}{T_\HH}-\frac{E_{\RR_k}-\mu_{\CC_k}}{T_\CC}=0,
    \end{aligned}&1\le k\le N,\\
    \\
    \begin{aligned}[b]
    &\frac{E_{\RR_k}-\mu_{\CC_k}}{T_\CC}-\frac{E_{\RR_k}-\mu_{\HH_k}}{T_\HH}\\
    &+\frac{E_{\LL_{k+1}}-\mu_{\CC_k}}{T_\CC}-\frac{E_{\LL_{k+1}}-\mu_{\HH_{k+1}}}{T_\HH}=0,
    \end{aligned}&1\le k<N.
\end{cases}
\end{equation}
To solve this system, we start by eliminating $\mu_{\HH_k}$ using the first line above,
\begin{equation}
    \mu_{\HH_k}=\frac{T_\HH}{2T_\CC}(\mu_{\CC_k}+\mu_{\CC_{k-1}})-\left(\frac{T_\HH}{T_\CC}-1\right)\frac{E_{\LL_k}+E_{\RR_k}}2.
\label{muH_fctmuC}
\end{equation}
For $k=1,\dots,N-1$, we then have
\begin{equation}
    \mu_{\CC_{k+1}}-2\mu_{\CC_k}+\mu_{\CC_{k-1}}=\left(1-\frac{T_\CC}{T_\HH}\right)(\Delta_{k+1}-\Delta_k),
\end{equation}
where we have defined $\Delta_k=E_{\RR_k}-E_{\LL_k}$. Defining $m_k=\mu_{\CC_k}-\mu_{\CC_{k-1}}$, the above equation becomes
\begin{equation}
   m_{k+1}=m_k+\left(1-\frac{T_\CC}{T_\HH}\right)(\Delta_{k+1}-\Delta_k).
\end{equation}
We can then readily obtain an explicit expression for $m_k$,
\begin{equation}
    m_k=\mu_{\CC_1}+\left(1-\frac{T_\CC}{T_\HH}\right)\sum_{l=1}^{k-1}(\Delta_{l+1}-\Delta_l),
\end{equation}
where we have used the fact that $m_1=\mu_{\CC_1}$ because $\mu_{\CC_0}=\mu_\LL=0$.
It is now straightforward to obtain $\mu_{\CC_k}$ as a function of $\mu_{\CC_1}$,
\begin{equation}
    \mu_{\CC_k}=\sum_{l=1}^km_k=k\mu_{\CC_1}+\left(1-\frac{T_\CC}{T_\HH}\right)\left(\sum_{l=1}^k\Delta_l-k\Delta_1\right).
\end{equation}
We deduce $\mu_{\CC_1}$ from the boundary condition~$\mu_{\CC_N}=\mu_\RR=0$,
\begin{equation}
    N\mu_{\CC_1}+\left(1-\frac{T_\CC}{T_\HH}\right)\left(\sum_{l=1}^N\Delta_l-N\Delta_1\right)=0,
\end{equation}
which yields
\begin{equation}
    \mu_{\CC_1}=\left(1-\frac{T_\CC}{T_\HH}\right)\left(\Delta_1-\frac1N\sum_{l=1}^N\Delta_l\right).
\end{equation}
We consequently have
\begin{equation}
    \mu_{\CC_k}=\left(1-\frac{T_\CC}{T_\HH}\right)\left(\sum_{l=1}^k\Delta_l-\frac kN\sum_{l=1}^N\Delta_l\right).
\end{equation}
Replacing $\mu_{\CC_k}$ by the above expression in Eq.~\eqref{muH_fctmuC}, we obtain
\begin{equation}
    \mu_{\HH_k}=\left(\frac{T_\HH}{T_\CC}-1\right)\left(\sum_{l=1}^k\Delta_l-\frac{2k-1}{2N}\sum_{l=1}^N\Delta_l-E_{\RR_k}\right).
\end{equation}

\section{Solution to the thermal control problem}
\label{app_ThCtrl}

In the linear response regime, the heat currents across the chain are given in Eqs.~\eqref{QHCk},
\begin{equation}
\begin{cases}
   \dot Q_{\HH_k}=\Gamma\Delta(E_{\RR_k}-E_{\LL_k}),&1\le k\le N,\\
   \dot Q_{\CC_k}=\Gamma\Delta(E_{\LL_{k+1}}-E_{\RR_k}),&1\le k<N,
\end{cases}
\label{sys_ThCtrl}
\end{equation}
where we have used the shorthand notation
\begin{equation}
    \Gamma=\frac\gamma{8N\hbar\kB\Theta}.
\end{equation}
Here, we are interested in the case where we are given a distribution of heat currents across the chain and seek to tune dot energies so as to realize this thermal distribution. This thermal control problem is then expressed by the system in Eqs.~\eqref{sys_ThCtrl} above, which must be solved for the dot energies~$E_{\LL_k},E_{\RR_k}$. Importantly, Eqs.~\eqref{sys_ThCtrl} do not define a linear system since~$\Delta$ contains all the dot energies~$E_{\LL_k},E_{\RR_k}$. However, we can linearize the system by noticing that
\begin{equation}
   \dot Q_\HH=\sum_{k=1}^N\dot Q_{\HH_k}=\Gamma\Delta^2.
\end{equation}
We can then write $\Delta=\pm\sqrt{\dot Q_{\HH}/\Gamma}$ since the second law of thermodynamics in Eq.~\eqref{2ndLaw} imposes $\dot Q_\HH\ge0$. The $\pm$~sign appearing here is a consequence of the fact that the system in Eq.~\eqref{sys_ThCtrl} is quadratic in the dot energies and thus admits two distinct solutions. It then becomes
\begin{equation}
    \begin{cases}
        E_{\RR_k}-E_{\LL_k}=\pm\dot Q_{\HH_k}/\sqrt{\Gamma\dot Q_\HH},&1\le k\le N,\\
        E_{\LL_{k+1}}-E_{\RR_k}=\pm\dot Q_{\CC_k}/\sqrt{\Gamma\dot Q_\HH},&1\le k<N.
    \end{cases}
\end{equation}
The first line above yields $E_{\RR_k}=E_{\LL_k}\pm\dot Q_{\HH_k}/\sqrt{\Gamma\dot Q_\HH}$, which leads to the following recurrence relation
\begin{equation}
    E_{\LL_{k+1}}=E_{\LL_k}\pm\frac{\dot Q_{\HH_k}+\dot Q_{\CC_k}}{\sqrt{\Gamma\dot Q_\HH}}.
\end{equation}
For $k=2,\dots,N$, we then have
\begin{equation}
    E_{\LL_k}=E_{\LL_1}\pm\sum_{l=1}^{k-1}\frac{\dot Q_{\HH_l}+\dot Q_{\CC_l}}{\sqrt{\Gamma\dot Q_\HH}}.
\end{equation}
$E_{\LL_1}$ is determined unambiguously if the value of the heat current out of the left lead is known. Indeed, Eq.~\eqref{QL} yields
\begin{equation}
    E_{\LL_1}=\pm\frac{\dot Q_\LL}{\sqrt{\Gamma\dot Q_\HH}}.
\end{equation}
We consequently obtain
\begin{subequations}
\begin{align}
    &E_{\LL_k}=\pm\frac1{\sqrt{\Gamma\dot Q_\HH}}\left(\dot Q_\LL+\sum_{l=1}^{k-1}(\dot Q_{\HH_l}+\dot Q_{\CC_l})\right),\\
    &E_{\RR_k}=\pm\frac1{\sqrt{\Gamma\dot Q_\HH}}\left(\dot Q_\LL+\sum_{l=1}^k\dot Q_{\HH_l}+\sum_{l=1}^{k-1}\dot Q_{\CC_l}\right).\label{ER_ThCtrl}
\end{align}
\end{subequations}
Interestingly, $E_{\RR_N}$ can alternatively be deduced from Eq.~\eqref{QR}:
\begin{equation}
E_{\RR_N}=\mp\frac{\dot Q_\RR}{\sqrt{\Gamma\dot Q_\HH}}.    
\end{equation}
This expression coincides with the result in Eq.~\eqref{ER_ThCtrl} since the first law of thermodynamics in Eq.~\eqref{1stLaw} imposes
\begin{equation}
   \dot Q_\LL+\sum_{k=1}^N\dot Q_{\HH_k}+\sum_{k=1}^{N-1}\dot Q_{\CC_k}=-\dot Q_\RR.
\end{equation}

\section{Validity of the thermal control solution}
\label{app_ValidityThCtrl}

The solution to the absolute thermal control problem in Eqs.~\eqref{ThCtrl_Solution} has been derived within the linear response regime. To verify the validity of this solution, it is necessary to ensure that the dot energies and cavity chemical potentials thus obtained allow for the expansion of the Fermi factors to first order. According to Eqs.~\eqref{muHCk} and~\eqref{ThCtrl_Solution}, the chemical potentials corresponding to the thermal control solution are
\begin{subequations}
\begin{align}
    &\mu_{\HH_k}=-\sqrt{\frac{8\hbar\kB T_\HH^2}{\gamma\Theta\langle\dot Q_\HH\rangle}}\left(\dot Q_\LL+\sum_{l=1}^{k-1}\dot Q_{\CC_l}+\left(k-\frac12\right)\langle\dot Q_\HH\rangle\right),\\
    &\mu_{\CC_k}=\sqrt{\frac{8\hbar\kB T_\CC^2}{\gamma\Theta\langle\dot Q_\HH\rangle}}\left(\sum_{l=1}^k\dot Q_{\HH_k}-k\langle\dot Q_\HH\rangle\right).
\end{align}
\end{subequations}
From Eq.~\eqref{j_WeakCoupling}, we understand that the linear response approximation can be applied to the current flowing through dot~$\LL_k$ if $|E_{\LL_k}-\mu_{\CC_{k-1}}|\ll\kB T_\CC$ and $|E_{\LL_k}-\mu_{\HH_k}|\ll\kB T_\HH$. We have similar conditions for the current flowing through dot~$\RR_k$, $|E_{\RR_k}-\mu_{\HH_k}|\ll\kB T_\HH$ and $|E_{\RR_k}-\mu_{\CC_k}|\ll\kB T_\CC$. To ensure that these properties hold all along the chain, it is simpler to analyze the sufficient conditions obtained using the triangle inequality, $|E_{\LL_k}|\ll\kB T_\CC$, $|E_{\RR_k}|\ll\kB T_\CC$, $|\mu_{\HH_k}|\ll\kB T_\HH$ and $|\mu_{\CC_k}|\ll\kB T_\CC$. Using the thermal control solution in Eqs.~\eqref{ThCtrl_Solution}, these straightforwardly reduce to
\begin{subequations}
\begin{align}
    &\frac1{\sqrt{\langle\dot Q_\HH\rangle}}\left|\dot Q_\LL+\sum_{l=1}^{k-1}(\dot Q_{\HH_l}+\dot Q_{\CC_l})\right|\ll\sqrt{\frac{\gamma\kB T_\CC^2}{\hbar\Theta}},\label{Validity_a}\\
    &\frac1{\sqrt{\langle\dot Q_\HH\rangle}}\left|\dot Q_\LL+\sum_{l=1}^k\dot Q_{\HH_l}+\sum_{l=1}^{k-1}\dot Q_{\CC_l}\right|\ll\sqrt{\frac{\gamma\kB T_\CC^2}{\hbar\Theta}},\label{Validity_b}\\
    &\frac1{\sqrt{\langle\dot Q_\HH\rangle}}\left|\dot Q_\LL+\sum_{l=1}^{k-1}\dot Q_{\CC_l}+\left(k-\frac12\right)\langle\dot Q_\HH\rangle\right|\ll\sqrt{\frac{\gamma\kB\Theta}\hbar},\label{Validity_c}\\
    &\frac1{\sqrt{\langle\dot Q_\HH\rangle}}\left|\sum_{l=1}^k\dot Q_{\HH_l}-k\langle\dot Q_\HH\rangle\right|\ll\sqrt{\frac{\gamma\kB\Theta}\hbar}.\label{Validity_d}
\end{align}
\label{Validity}%
\end{subequations}


We now adopt a recursive approach to draw general conclusions about the distributions for the heat currents~$\dot Q_{\HH_k}$ and~$\dot Q_{\CC_k}$ from these conditions. We then assume that the conditions in Eqs.~\eqref{Validity} are satisfied at site~$k-1$, that is,
\begin{subequations}
\begin{align}
    &\frac1{\sqrt{\langle\dot Q_\HH\rangle}}\left|\dot Q_\LL+\sum_{l=1}^{k-2}(\dot Q_{\HH_l}+\dot Q_{\CC_l})\right|\ll\sqrt{\frac{\gamma\kB T_\CC^2}{\hbar\Theta}},\\
    &\frac1{\sqrt{\langle\dot Q_\HH\rangle}}\left|\dot Q_\LL+\sum_{l=1}^{k-1}\dot Q_{\HH_l}+\sum_{l=1}^{k-2}\dot Q_{\CC_l}\right|\ll\sqrt{\frac{\gamma\kB T_\CC^2}{\hbar\Theta}},\label{Validity_b-k-1}\\
    &\frac1{\sqrt{\langle\dot Q_\HH\rangle}}\left|\dot Q_\LL+\sum_{l=1}^{k-2}\dot Q_{\CC_l}+\left(k-\frac32\right)\langle\dot Q_\HH\rangle\right|\ll\sqrt{\frac{\gamma\kB\Theta}\hbar},\label{Validity_c-k-1}\\
    &\frac1{\sqrt{\langle\dot Q_\HH\rangle}}\left|\sum_{l=1}^{k-1}\dot Q_{\HH_l}-(k-1)\langle\dot Q_\HH\rangle\right|\ll\sqrt{\frac{\gamma\kB\Theta}\hbar}.\label{Validity_d-k-1}
\end{align}
\end{subequations}
We now look for the constraints to impose on heat currents~$\dot Q_{\CC_{k-1}}$ and~$\dot Q_{\HH_k}$ so that the same conditions hold at site~$k$. From Eq.~\eqref{Validity_b-k-1}, we see that Eqs.~\eqref{Validity_a} and~\eqref{Validity_b} will be satisfied at site~$k$ if we choose $\dot Q_{\CC_{k-1}}$ and $\dot Q_{\HH_k}$ such that
\begin{align}
    &\frac{\left|\dot Q_{\CC_{k-1}}\right|}{\sqrt{\langle\dot Q_\HH\rangle}}\ll\sqrt{\frac{\gamma\kB T_\CC^2}{\hbar\Theta}},\\
    &\frac1{\sqrt{\langle\dot Q_\HH\rangle}}\left|\dot Q_{\CC_{k-1}}+\dot Q_{\HH_k}\right|\ll\sqrt{\frac{\gamma\kB T_\CC^2}{\hbar\Theta}}.
\end{align}
Similarly, taking into account Eqs.~\eqref{Validity_c-k-1} and~\eqref{Validity_d-k-1}, we find that Eqs.~\eqref{Validity_c} and~\eqref{Validity_d} are satisfied at site~$k$ if
\begin{align}
    &\frac1{\sqrt{\langle\dot Q_\HH\rangle}}\left|\dot Q_{\CC_{k-1}}+\langle\dot Q_\HH\rangle\right|\ll\sqrt{\frac{\gamma\kB\Theta}\hbar},\\
    &\frac1{\sqrt{\langle\dot Q_\HH\rangle}}\left|\dot Q_{\HH_k}-\langle\dot Q_\HH\rangle\right|\ll\sqrt{\frac{\gamma\kB\Theta}\hbar}.
\end{align}

Our recursive approach then demonstrates that the conditions in Eqs.~\eqref{Validity} should be satisfied for any site in the chain if the following constraints are imposed on all heat currents:
\begin{subequations}
\begin{align}
    &\frac{\left|\dot Q_{\CC_k}\right|}{\sqrt{\langle\dot Q_\HH\rangle}}\ll\sqrt{\frac{\gamma\kB T_\CC^2}{\hbar\Theta}},\\
    &\frac1{\sqrt{\langle\dot Q_\HH\rangle}}\left|\dot Q_{\CC_{k-1}}+\dot Q_{\HH_k}\right|\ll\sqrt{\frac{\gamma\kB T_\CC^2}{\hbar\Theta}},\label{Validity_b-k}\\
    &\frac1{\sqrt{\langle\dot Q_\HH\rangle}}\left|\dot Q_{\CC_k}+\langle\dot Q_\HH\rangle\right|\ll\sqrt{\frac{\gamma\kB\Theta}\hbar},\label{Validity_c-k}\\
    &\frac1{\sqrt{\langle\dot Q_\HH\rangle}}\left|\dot Q_{\HH_k}-\langle\dot Q_\HH\rangle\right|\ll\sqrt{\frac{\gamma\kB\Theta}\hbar}.\label{Validity_d-k}
\end{align}
\label{Validity-k}%
\end{subequations}
Note that Eqs.~\eqref{Validity_c-k} and~\eqref{Validity_d-k} are redundant since, according to Eq.~\eqref{Validity_b-k}, the distributions for $\dot Q_{\HH_k}$ and $\dot Q_{\CC_k}$ should be almost opposite to one another. As such, a constraint imposed on the dispersion of one of these distribution will necessarily apply to the other one as well. Typically, the local conditions in Eq.~\eqref{Validity-k} will be satisfied if we choose a thermal distribution such that
\begin{subequations}
\begin{align}
    &\langle\dot Q_\HH\rangle\ll\frac{\gamma\kB T_\CC^2}{\hbar\Theta},\\
    &\frac{\delta^2}{\langle\dot Q_\HH\rangle}\ll\frac{\gamma\kB T_\CC^2}{\hbar\Theta},\\
    &\frac{\sigma^2}{\langle\dot Q_\HH\rangle}\ll\frac{\gamma\kB\Theta}\hbar,
\end{align}
\end{subequations}
where we have defined
\begin{align}
    &\delta^2=\frac1N\sum_{k=1}^N\left(\dot Q_{\CC_{k-1}}+\dot Q_{\HH_k}\right)^2,\\
    &\sigma^2=\frac1N\sum_{k=1}^N\left(\dot Q_{\HH_k}-\langle\dot Q_\HH\rangle\right)^2.
\end{align}

\bibliography{ref}

\end{document}